\documentclass[5p,sort&compress]{elsarticle}

\usepackage{lineno}
\usepackage{graphics}
\usepackage[version=3]{mhchem} % Formula subscripts using \ce{}
\usepackage{siunitx} % For easy use of units
\usepackage{url,subfigure,color,booktabs}
\usepackage[hidelinks]{hyperref} % clickable ref's etc, with no ugly boxes
\modulolinenumbers[5]

\usepackage{fancyhdr}
\begin{document}

\begin{frontmatter}

\title{Liquid Marble Interaction Gate\\ for Collision-Based Computing}

\author[UWE]{Thomas C.\ Draper}
\author[UWE]{Claire Fullarton}
\author[UWE]{Neil Phillips}
\author[bio]{Ben~P.~J.~De~Lacy~Costello}
\author[UWE]{Andrew Adamatzky\corref{mycorrespondingauthor}}

\cortext[mycorrespondingauthor]{Corresponding author}
\ead{Andrew.Adamatzky@uwe.ac.uk}

\address[UWE]{Unconventional Computing Laboratory, \\University of the West of England, Bristol, United Kingdom}
\address[bio]{Institute of Biosensing Technology, Centre for Research in Biosciences,\\ University of the West of England, Bristol, United Kingdom}

\begin{abstract}
Liquid marbles are microlitre droplets of liquid, encapsulated by self-organised hydrophobic particles at the liquid/air interface. They offer an efficient approach for manipulating liquid droplets and compartmentalising reactions in droplets. Digital fluidic devices employing liquid marbles might benefit from having embedded computing circuits without electronics and moving mechanical parts (apart from the marbles). We present an experimental implementation of a collision gate with liquid marbles. Mechanics of the gate follows principles of Margolus' soft-sphere collision gate. Boolean values of the inputs are given by the absence ({\sc False}) or presence ({\sc True}) of a liquid marble. There are three outputs: two outputs are trajectories of undisturbed marbles (they only report {\sc True} when just one marble is present at one of the inputs), one output is represented by trajectories of colliding marbles (when two marbles collide they lose their horizontal momentum and fall), this output reports {\sc True} only when two marbles are present at inputs. Thus the gate implements {\sc AND} and {\sc AND-NOT} logical functions. We speculate that by merging trajectories representing {\sc AND-NOT} output into a single channel one can produce a one-bit half-adder. Potential design of a one-bit full-adder is discussed, and the synthesis of both a pure nickel metal and hybrid nickel/polymer liquid marble is reported.
\end{abstract}


\end{frontmatter}

%\linenumbers

\section{Introduction}

Since their inception in 2001 \cite{Aussillous2001}, liquid marbles (LMs) have been a source of growing interest across fields as diverse as medicine \cite{Sarvi2015,Ledda2016}, engineering \cite{Fujii2017,Tian2010a} and chemistry \cite{Paven2016,Wei2016}. LMs are constructed from microlitre droplets of water, supported by a layer of hydrophobic particles on the surface. In this manner, the hydrophobic particles minimise the comparatively high surface energy of water by encapsulating the droplet, and keeping it near spherical. This permits the water droplet to remain non-wetting on many (traditionally wettable) surfaces. 

There are two major variables affecting the properties of a LM: the core and the coating. Traditionally the encapsulated liquid is water, although there are a range of both common (glycerol \cite{Aussillous2001}) and uncommon (petroleum \cite{Bormashenko2015a}) alternatives. The coating provides the largest affect on the mechanical properties of the marble, as it is the coating that interacts with the surface the LM is resting on. Coating parameters that can be modified to impart the desired properties include the composition, grain size, and mix ratio. By varying these, it is possible to adapt a liquid marble to many different situations.

Liquid marbles have previously been investigated as fluidic transport devices. They are ideally suited to the transport of microlitre quantities of liquid, due to their non-wetting nature. Recent progress have been made in this area, with LM movement initiated by magnets \cite{Khaw2016}, electrostatic fields \cite{Aussillous2006}, gravity \cite{Aussillous2001}, lasers \cite{Paven2016} and the Maran\-goni effect \cite{Ooi2015} all reported. The movement of LMs can be exploited for chemical reactions, by controlling the time and place of reagent mixing. This has been demonstrated both with the coalescing of LMs \cite{Chen2017}, and in the controlled destruction of LMs once they arrive at a chosen location \cite{dupin2009stimulus,Fujii2010}.

In interaction gates, Boolean values of inputs and outputs are represented by presence of physical objects at given site at a given time. If an object is present at input/output we assume that logical value of the input/output is {\sc True}; if the object is absent the logical value is {\sc False}. The signal-object realise a logical function when they pass through a collision site. The objects might fuse, annihilate or deflect on impact. 

\begin{figure}[htbp]
\centering
\subfigure[]{\includegraphics[scale=0.3]{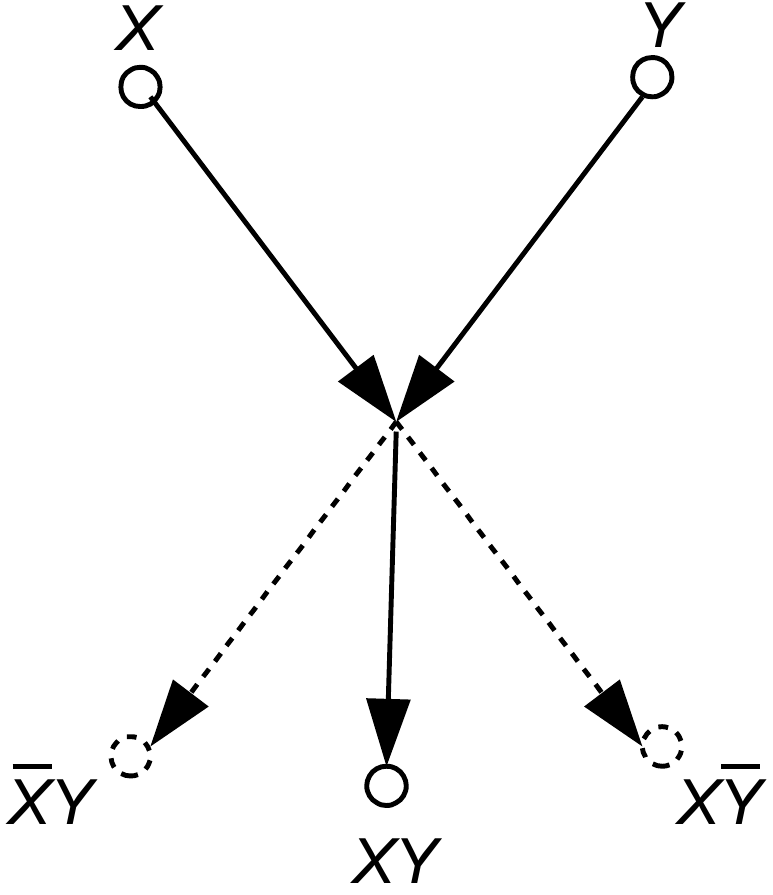}}
\subfigure[]{\includegraphics[scale=0.3]{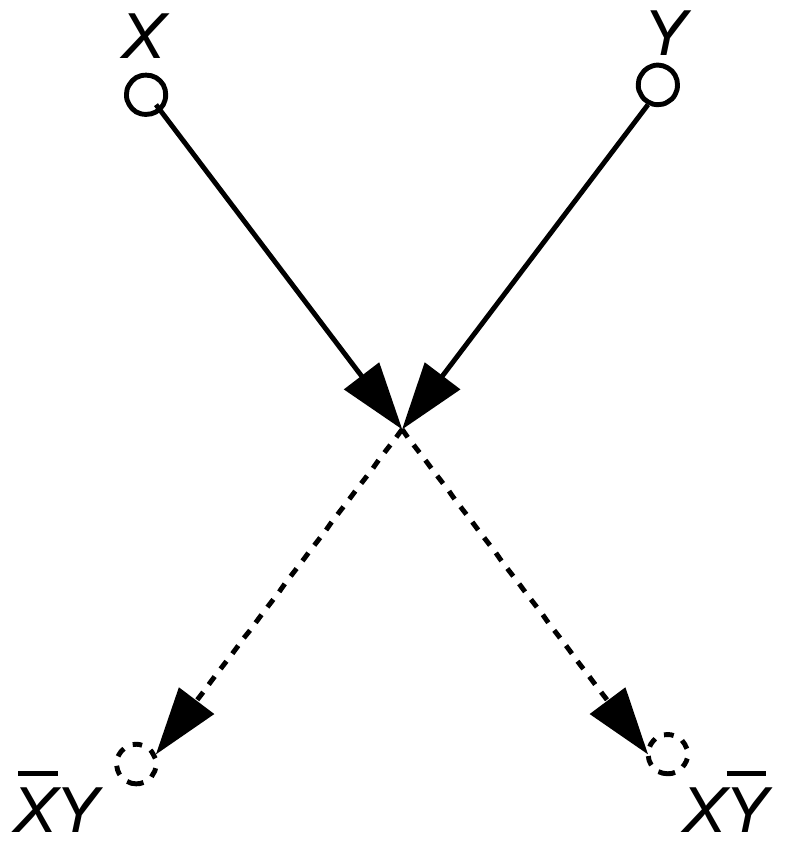}}\\
\subfigure[]{\includegraphics[scale=0.3]{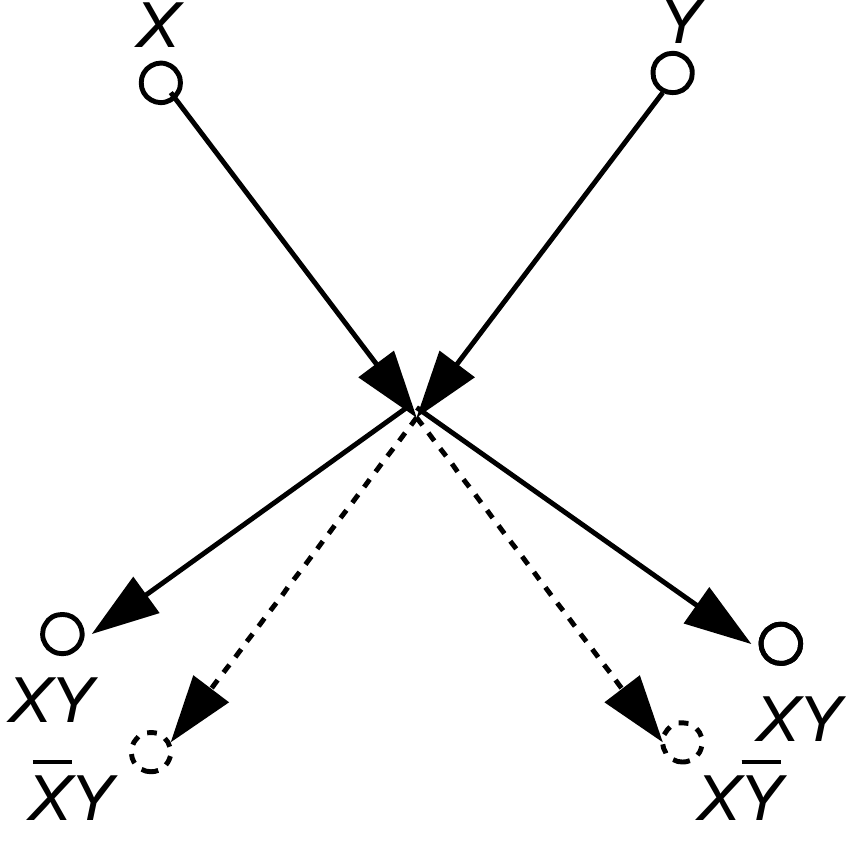}}
\subfigure[]{\includegraphics[scale=0.3]{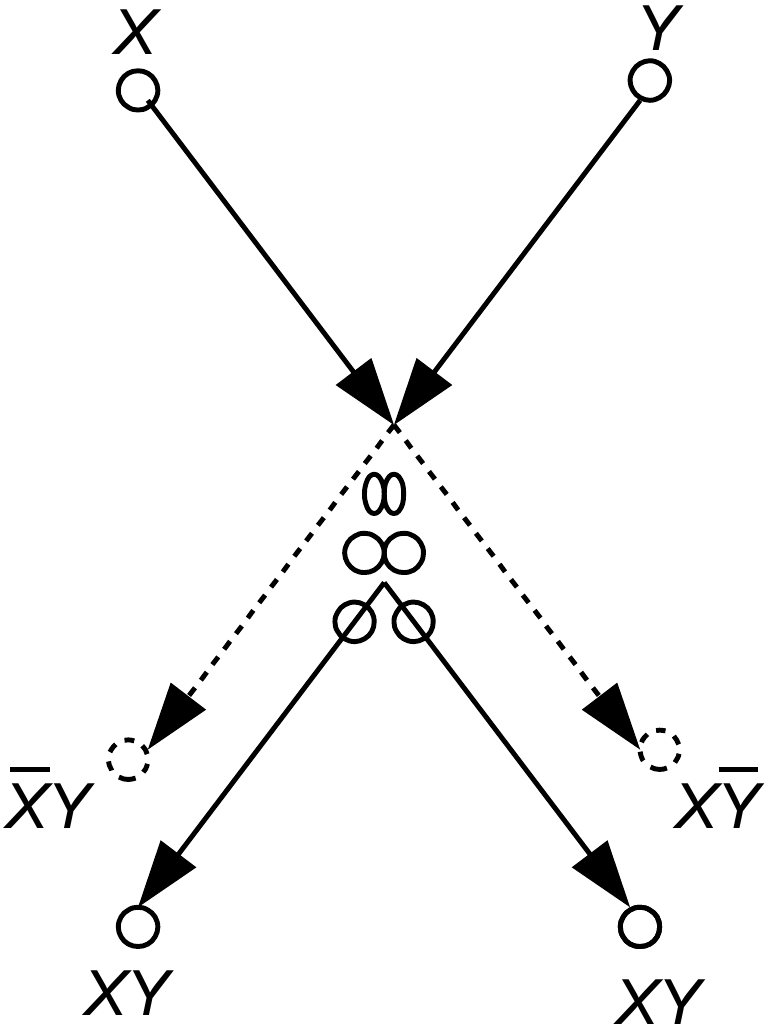}}
\caption{Interaction-based gates. 
(a)~Fusion of signals.
(b)~Annihilation of signals.
(c)~Elastic deflection of signals.
(d)~Non-elastic deflection of signals.}
\label{fig:gate}
\end{figure}

The fusion gate (figure~\ref{fig:gate}(a)) was first implemented in fluidic devices in the 1960s. The  gate is the most well known (on a par with the bistable amplifier) device in fluidics~\cite{peter1965and, belsterling1971fluidic}: two nozzles are placed at right angles to each other, when there are jet flows in both nozzle they collide and merge  into a single jet entering the central outlet.  If the jet flow is present only in one of the input nozzles it goes into the vent. The central outlet represents {\sc and} and the vent represent {\sc and-not}. The fusion-based gate was also employed in designs of computing circuits in Belousov-Zhabotinsky medium \cite{adamatzky2004collision, adamatzky2007binary, toth2010simple, adamatzky2011towards}, where excitation wave-fragments merge when collide; in the actions of slime mould \cite{Mayne2015Vesicles}, when distributing vesicles collide; and a crab-based gate \cite{gunji2011robust}, were swarms of solider crabs merge into a single swarm.

In the annihilation gate (figure~\ref{fig:gate}b) signals disappear on impact. This gates has two-inputs and two-outputs, each of the outputs represents {\sc and-not}. Computational universality of the Conway's Game of Life cellular automata was demonstrated using annihilation based collisions between gliders~\cite{berlekamp1982winning}. We can also implement the annihilation gate by colliding excitation wave-fragments at certain angles~\cite{adamatzky2011polymorphic}. 

Key deficiency of the fusion and annihilation gates is that, when implemented in media other than excitable spatially-extended systems, they do not preserve physical quantity of signals, e.g. when two signals merge the output signal will have a double mass of a signal input signal. This deficiency is overcome in the conservative logic, proposed by Fredkin and Toffoli in 1978~\cite{fredkin2002conservative}. The logical value are represented by solid elastic bodies, aka billiard balls, which deflect when made to collide with one another (figure~\ref{fig:gate}c). Intact output trajectories of the balls represent {\sc and-not} function, output trajectories of deflected balls represent {\sc and} function. The gates based on elastic collision led to development of a reversible (both logically and physically) gate:  Fredkin~\cite{fredkin2002conservative} and Toffoli~\cite{toffoli1980reversible} gates, which are the key elements of low-power  computing circuits \cite{de2011reversible, bennett1988notes}, and amongst the key components of quantum~\cite{berman1994quantum, barenco1995elementary, smolin1996five, zheng2013implementation} and optical~\cite{kostinski2009experimental} computing circuits. 

The soft-sphere collision gate proposed by Margolus~\cite{margolus2002universal} gives us a rather realistic representation of interaction gates with real-life physical and biological bodies (figure~\ref{fig:gate}d). Logical value $x=1$ is given by a ball presented in input trajectory marked $x$, and $x=0$ by the absence of the ball in the input trajectory $x$; the same applies to $y=1$ and $y=0$, respectively. When the two balls, approaching the collision gate along paths $x$ and $y$ collide, {they} compress but then spring back and reflect. As a result, the balls come out along the paths marked $xy$. If only one ball approaches the gate, that is for inputs $x=1$ and $y=0$ or $x=0$ and $y=1$, the balls exit the gate via path $x\overline{y}$ (for input $x=1$ and $y=0$) or $\overline{x}y$ (for input $x=0$ and $y=1$). Soft-sphere-like gates have  been implemented using microlitre sized water droplets on a superhydrophobic copper surface \cite{Mertaniemi2012}. Using channels cut into the surface, {\sc not-fanout}, {\sc and-or} and {\sc flip-flop} gates were demonstrated. The water droplets only rebounded in a very narrow collision property window, and a thorough \& complete superhydrophobic surface treatment was required.

In this paper, we report the first exploitation of liquid marbles for implementation of interaction gates. The gate realised in the experimental laboratory conditions is a combination of the fusion and the Margolus gate: output trajectories of collided liquid marbles are so close that they can be interpreted as a single output. That said, if required, two liquid marbles at the output can be diverted along different paths to conserve a number of signals. By taking advantage of the liquid marbles' inherently low hysteresis, high tunability and capacity for enhanced versatility, we demonstrate the first step toward liquid marble facilitated, collision-based computing.

\section{Materials \& Method}
\subsection{Regular \& Reliable Liquid Marble Formation}

We first developed a technique for the regular and automatic formation of invariable LMs. This was achieved by programming a syringe driver (CareFusion Alaris GH) to feed a 21 gauge needle (\SI{0.8}{\milli\metre} diameter) at a typical rate of \SI{7.0}{\milli\litre\per\hour}. The rate can be easily increased or decreased, and this rate gave sufficiently fast LM formation for our purpose. The produced droplets (\SI[separate-uncertainty = true,multi-part-units = single]{11.6 +- 0.16}{\micro\litre}) were permitted to fall onto a sheet of acrylic, slanted to \ang{20} from horizontal, and surface-treated with a commercial hydrophobic spray (Rust-Oleum\textsuperscript{\textregistered} NeverWet\textsuperscript{\textregistered}). This formed beads of water, which were allowed to roll over a bed of appropriate hydrophobic powder. The result was a continuous `stream' of LMs with the same volume, coating and coating thickness. It should be noted that whilst the forming of LMs by running droplets down a powder slope has been separately developed by another group \cite{Fujii2010}, our system prevents premature destruction of the powder bed by initially preforming the droplet on a treated hydrophobic surface.

\subsection{Maintaining Timing for Collisions}

In collision based computing, accurate timing is essential. As signals propagate through the system they must remain in sync, or the operation of many logic gates fails. In order to address this, an innovative system of electromagnets (EMs) was implemented. This was possible due to the generation of novel LMs with a mix of ultra-high density polyethylene (UHDPE) (Sigma-Aldrich, \numrange[range-phrase = --]{3}{6e6}~\si{\gram\per\mole}, grain size approximately \SI{100}{\micro\metre}) and nickel (GoodFellow Metals, \SI{99.8}{\percent}, grain size \SIrange{4}{7}{\micro\metre}). A typical Ni/UHDPE coating was \SI{2.5}{\mg}. The use of UHDPE provides strength and durability, and the inclusion of ferromagnetic nickel allows for a versatile magnetic LM.

By positioning an electromagnet (\SI{100}{\newton}, \SI{12.0}{\volt} DC, \SI{29 x 22}{\mm}) behind the acrylic slope, the rolling LM can be captured and released at will, by the switching on and off of said electromagnet. By controlling multiple, spatially-isolated electromagnets in series, non-concurrent LMs can be easily synchronised. To our knowledge, this is the first time electromagnets have been used to provide timing control with liquid marbles.

\subsection{Gate Design for Liquid Marble Collisions}

The collision gate was designed to allow for the colliding LMs to have a free path post-collision. This enabled the monitoring of the LM paths, and the future design and implementation of exiting pathways, creating a logic gate.

\begin{figure}[htbp]
  \centering
  \includegraphics[width=0.99\linewidth]{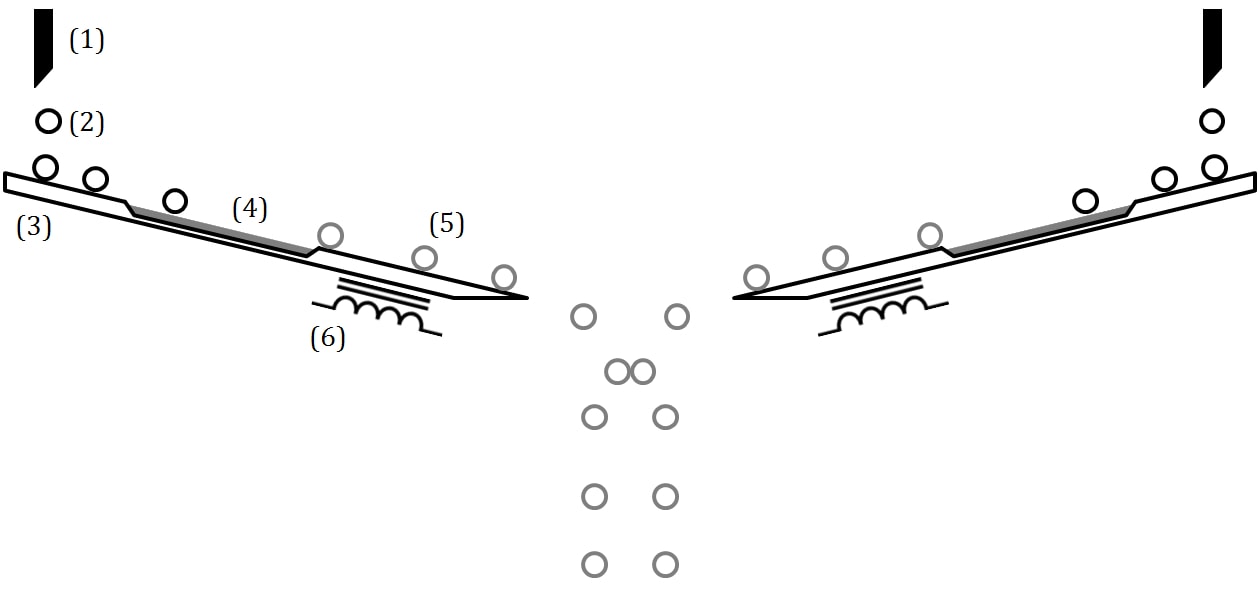}
  \caption{Schematic of our LM collider. Labelled numbers are: (1)~syringe needle, (2)~uncoated water droplet, (3)~acrylic ramp, (4)~hydrophobic powder bed, (5)~liquid marble, and (6)~electromagnet. Droplets form and fall out of the two syringe needles, landing on a superhydrophobic surface. They then roll over a bed of Ni/UHDPE powder, before being stopped and held stationary by the electromagnets. These electromagnets are then deactivated simultaneously, allowing the LMs to roll off and collide.}
  \label{fig:LM-collider}
\end{figure}

\begin{figure}[htbp]
\centering
\subfigure[]{\includegraphics[width=0.48\textwidth]{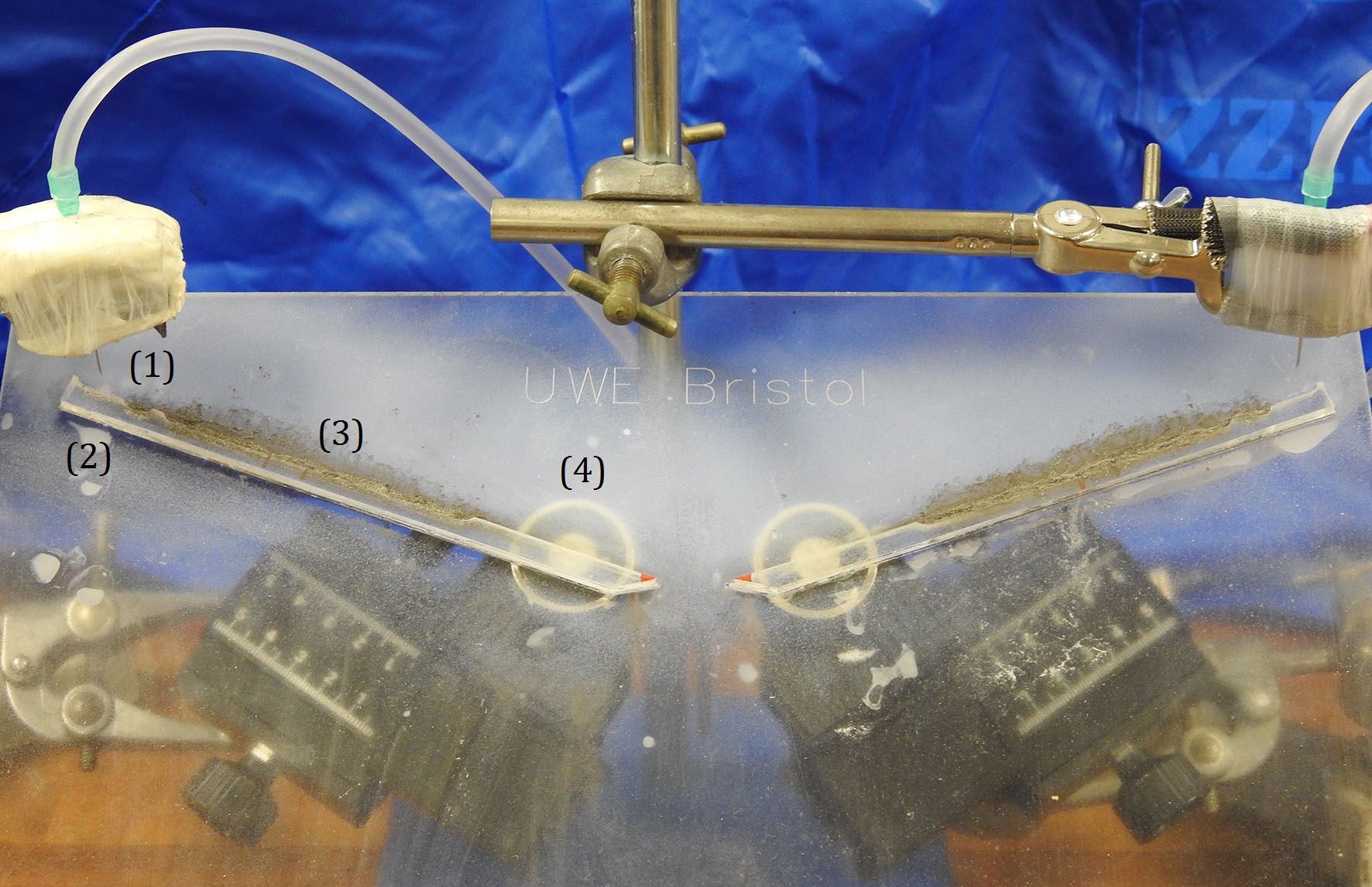}}
\subfigure[]{\includegraphics[width=0.23\textwidth]{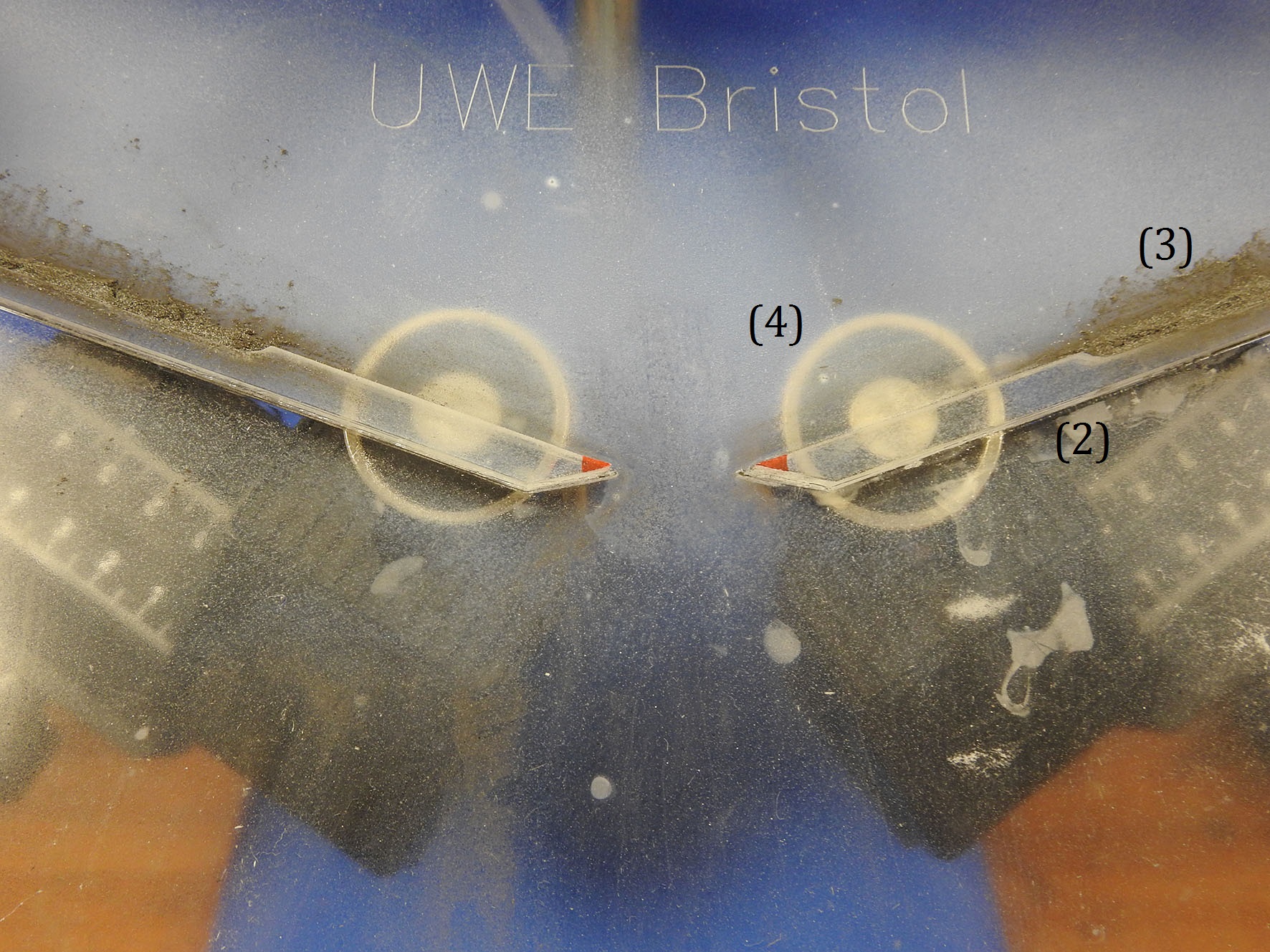}}
\subfigure[]{\includegraphics[width=0.23\textwidth]{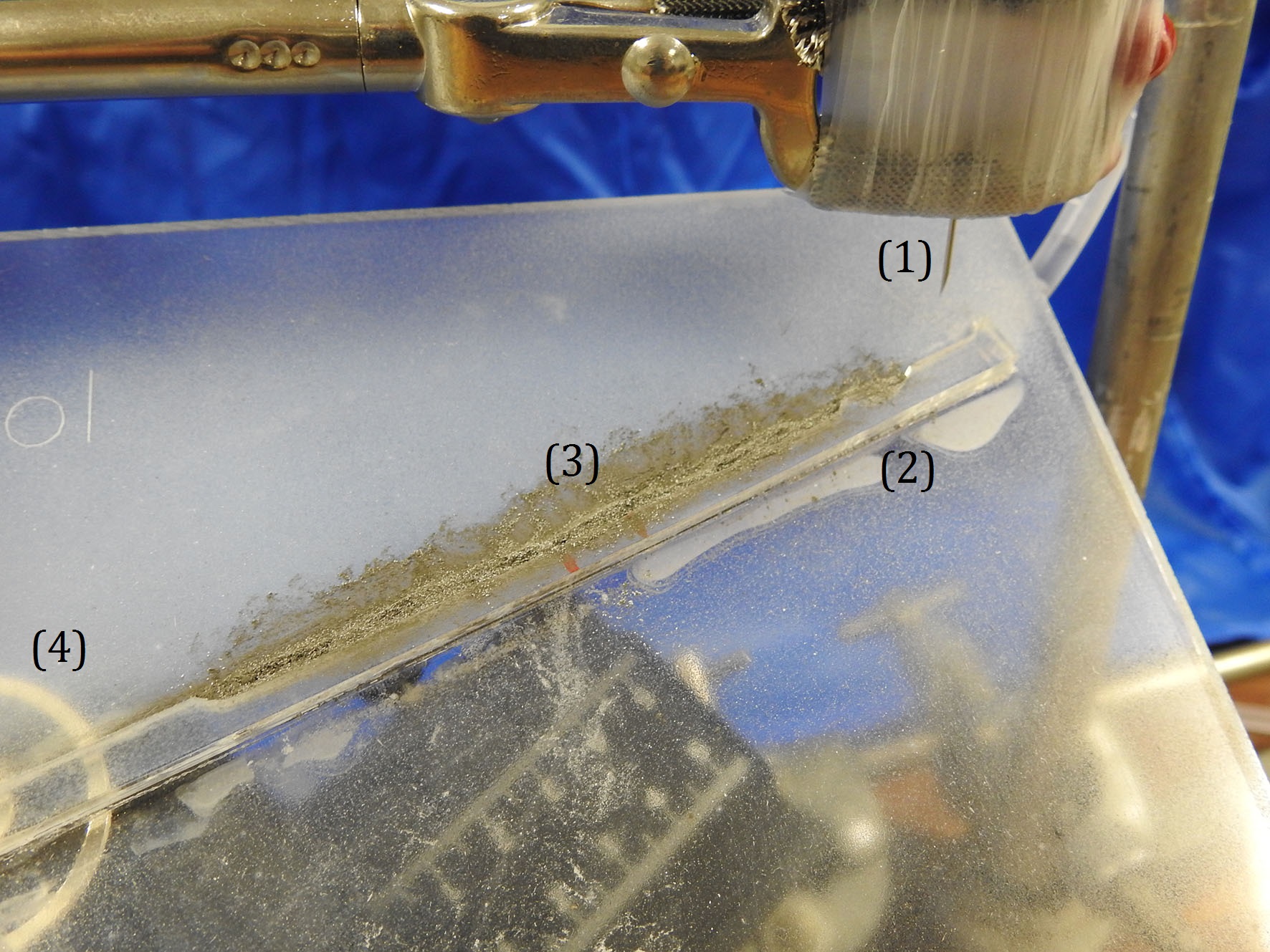}}
\caption{Photographs of our LM collider, showing 
(a)~the overall layout, 
(b)~an upclose of the magnetic breaking/release area, and
(c)~an upclose of the droplet formation area. Labels are: (1)~syringe needle, (2)~acrylic ramp, (3)~hydrophobic powder bed, and (4)~electromagnet.}
\label{fig:LM-collider-photo}
\end{figure}

Two \SI{16.0}{\centi\metre} acrylic pathways were slanted towards each other at \ang{20}, affixed to an acrylic base sheet (\SI{3.0}{\mm} thick). The acrylic base sheet was then aligned with a pitch of \ang{38} from horizontal, giving a final LM pathway slope of \ang{16} from the horizontal plane. This gave reliable LM rolling without extreme angles. The gap between the two slanted pathways was set at \SI{1.6}{\centi\metre}, after empirical testing. A \SI{2.0}{\centi\metre}, length at the top of each pathway was made hydrophobic, as discussed above. Parallel auto-formation of hybrid LMs was achieved using the syringe driver, delivering \SI{11.6}{\micro\litre} of water per syringe per drop. Each droplet of water was permitted to land on the treated section of each slope, before rolling across the powder beds of UHDPE and Ni to form LMs. The two rolling LMs were then captured using the electromagnets, allowing for any slight timing deviations to be accounted for. On controlled synchronous (or asynchronous) release of the electromagnets, the LMs simultaneously roll off the acrylic ramps on collision trajectories. Collisions were recorded at 120 fps using a Nikon Coolpix P900, and played back frame-by-frame for analysis. A schematic of our LM collider can be seen in figure \ref{fig:LM-collider}, and photographs of our LM collider can be seen in figure~\ref{fig:LM-collider-photo}.

\section{Discussion}
\subsection{Liquid Marble Lifetime}

\begin{figure}[htbp]
  \centering
  \subfigure[]{\includegraphics[width=0.4\linewidth]{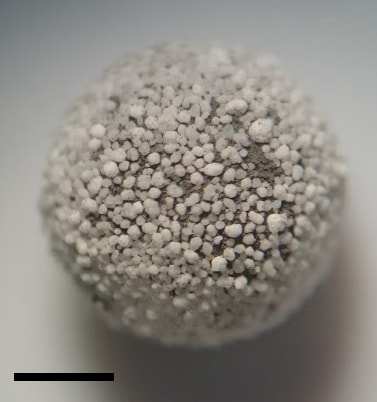}}
  \subfigure[]{\includegraphics[width=0.4\linewidth]{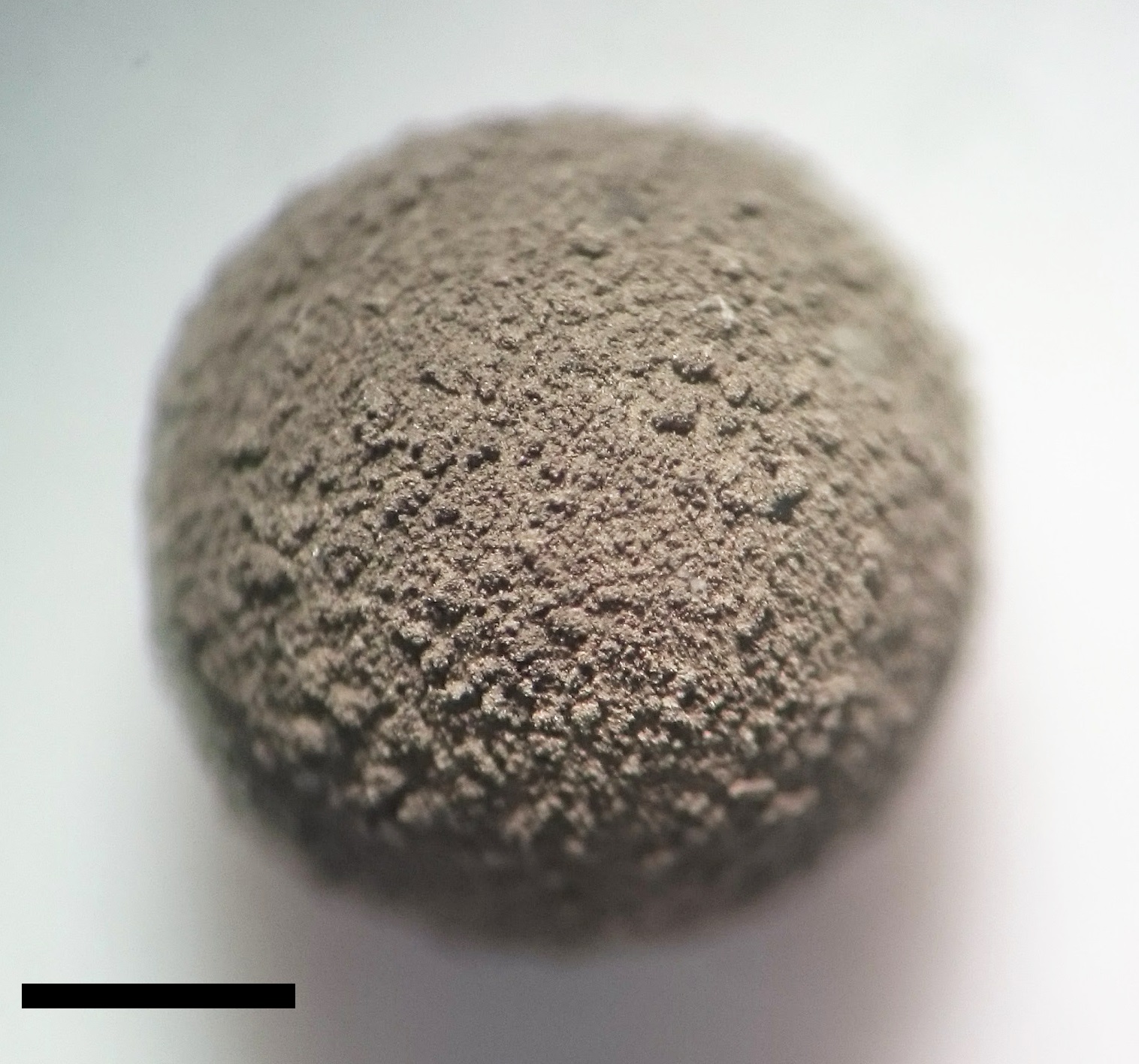}}
  \caption{Magnified photographs of (a) a Ni/UHDPE LM, nickel can clearly been seen between the UHDPE particles; and (b) a pure nickel LM.  Scale bars are \SI{1.0}{\mm}. }
  \label{fig:LM-photo}
\end{figure}

For a LM to be useful in a computing device, it has to have an appreciable lifetime. This is problematic for water based LM, as the gas permeability of LMs has previously been both established and exploited \cite{Eshtiaghi2009,Tian2010a,Tian2010,Tian2013}. As such, lifetime experiments were conducted on UHDPE LMs, Ni LMs, water droplets, and our new Ni/UHDPE hybrid LMs. Evaporation studies were conducted under ambient conditions, using \SI{10.0}{\micro\litre} of DI water. UHDPE LMs were generated by rolling a droplet of water on a powder bed of UHDPE. Nickel LMs were made using a superhydrophobic surface (see above) to pre-form the droplet sphere, before rolling in an appropriate powder. A magnified view of both a Ni/UHDPE hybrid LM and a nickel LM can be seen in figure~\ref{fig:LM-photo}.

From the data shown in table~\ref{tab:evap}, it can be seen that both the nickel and UHDPE LMs have slower evaporation rates than a pure water droplet. This is expected, as the solid particles on the surface of the liquid form an (incomplete) barrier to evaporation. It also supports previous studies \cite{Dandan2009}. Experiments also indicated that pure UHDPE LMs evaporate at a comparable rate to pure nickel LMs. This is due to a balancing act between the short narrow pores of the nickel LM, the long wide channels of the UHDPE LM, and the much larger contact angle of UHDPE compared to nickel \cite{Trevoy1958,Ammosova2015}. The larger grain size of the UHDPE creates longer channels for water vapour to traverse. This  results in a smaller water vapour concentration gradient. 

\begin{table}[htbp]
  \centering
  \caption{Comparison of the evaporation rates for \SI{10}{\micro\litre} LMs and an uncoated water droplet. }
    \begin{tabular}{cc}
    \toprule
    \textbf{LM Coating} & \textbf{Initial Evaporation Rate} \\
     & / \si{\mg\per\minute} \\
    \midrule
    (Water Droplet) & 0.1392 \\
    Ni    & 0.1133 \\
    UHDPE & 0.1107 \\
    Ni/UHDPE & 0.0998 \\
    \bottomrule
    \end{tabular}%
  \label{tab:evap}%
\end{table}%

It is noteworthy that the new hybrid LM offers the best protection, with the lowest rate of evaporation. It is suggested that this is due to differences in the nickel and UHDPE particle sizes. This difference is clearly visible in the magnified photograph shown in figure~\ref{fig:LM-photo}(a). The larger UHDPE particles offer good resistance to evaporation, due to their thickness. However, this also leads to large gaps between the particles, due to poor packing. This problem is alleviated by the nickel particles filling the available space. The use of two differently sized spheres for 3D spherical packing is well documented \cite{Yamada2011,Sohn1968}, and has been shown to increase the maximum perfect packing density beyond the \num{0.74} limit of a single-sized 3D sphere packing, to \num{0.93} for an ideal binary system \cite{Kansal2002}. In this instance, due to the multilayer nature of the LM coating \cite{dupin2009stimulus}, it is more appropriate to relate to 3D sphere packing than 2D circle packing.

\subsection{Liquid Marble Collisions}

If the gate timings are not accurate, then the LMs will continue on their separate paths and shall not collide. If the timings are accurate, and it is taken that the two LMs are identical, then there are three possible outcomes of the collision. Firstly, that the LMs collide with some elastic property, and then continue on two distinct and new paths. Secondly, that the LMs collide with no elastic property, then continue vertically as two adjacent, but distinct, LMs. Thirdly, that on collision the LMs coalesce into a larger single LM with zero lateral velocity, which continues vertically down.

\begin{figure}[htbp]
\centering
\subfigure[]{\includegraphics[width=0.35\textwidth]{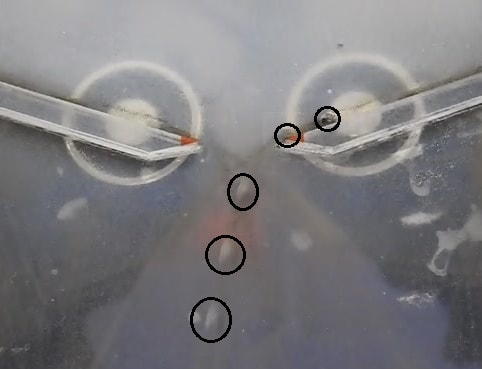}}
\subfigure[]{\includegraphics[width=0.35\textwidth]{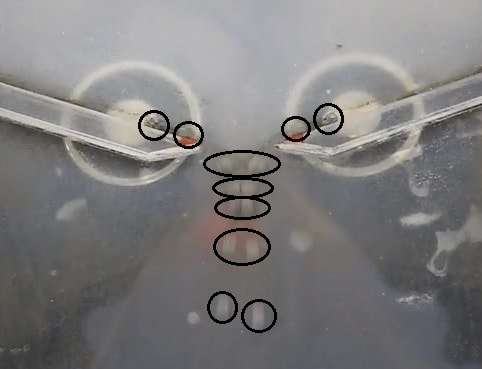}}
\caption{Overlaid still frames of (a) a single LM, with frames at \SIlist{0;142;209;242;267}{\milli\second}; 
and (b) two colliding LMs, with frames at \SIlist{0;125;200;217;225;250;275}{\milli\second}.}
\label{fig:single-bounce}
\end{figure}

\begin{figure}[htbp]
\centering
\subfigure[]{\includegraphics[width=0.4\linewidth]{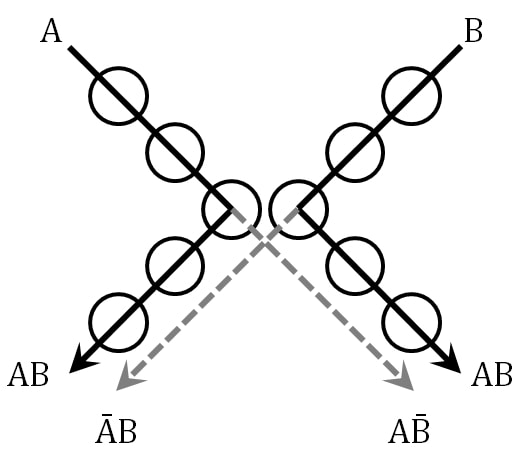}}
\subfigure[]{\includegraphics[width=0.4\linewidth]{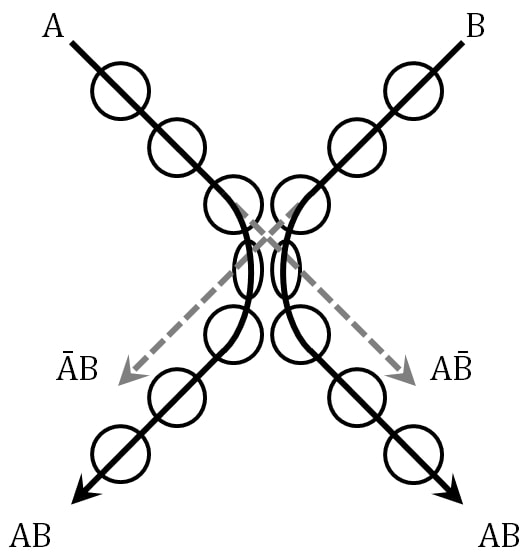}}
\subfigure[]{\includegraphics[width=0.7\linewidth]{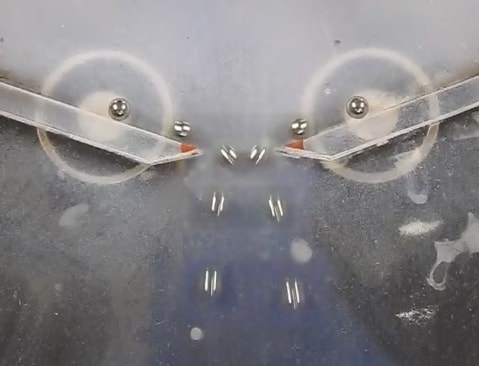}}
\caption{Showing the colliding and non-colliding routes for (a) BBM and (b) SSM pathways. (c) A collision of steel balls, following BBM under gravity (cf.\ SSM LMs above). Frame times are \SIlist{0;175;242;284;325}{\ms}.}
\label{fig:BBM-SSM}
\end{figure}

Video snapshots showing both a single uninterrupted and a colliding pair of LMs, can be seen in figure~\ref{fig:single-bounce}. Our experiments demonstrate that LMs collide in an elastic manner. This is unsurprising, due to their previously reported soft-shell and compressible nature \cite{Aussillous2001,Bormashenko2015}. It also supports the previously published, linear, non-coalescing collision of LMs \cite{Bormashenko2015}. By monitoring the collisions at \SI{120}{fps}, it was observed that LMs behave like two soft balls, acting in a manner described in the Soft Sphere Model (SSM), known as a Margolus gate \cite{margolus2002universal}. A video of a typical collision can be seen in the supporting information. The important distinction between the SSM and the better known Billiard Ball Model (BBM) \cite{fredkin2002conservative}, is the exit points of the colliding particles compared to the non-colliding particles. In the BBM, the particles are taken to be hard spheres, which instantly rebound off each other --- leading to the \textbf{AB} paths being outside the corresponding \textbf{\={A}B} and \textbf{A\={B}} paths. In contrast, the SSM accounts for the finite and appreciable amount of time required for real-world soft spheres to rebound. The result is that the AB paths move to lie inside the unchanged \textbf{\={A}B} and \textbf{A\={B}} paths. The BBM and SSM pathways can be seen in figures \ref{fig:BBM-SSM}(a) and \ref{fig:BBM-SSM}(b), respectively.

\begin{figure}[htbp]
  \centering
  \includegraphics[width=0.7\linewidth]{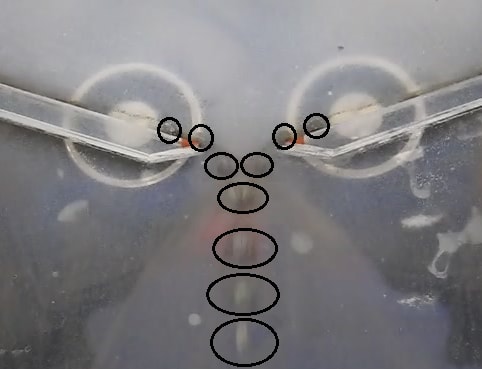}
  \caption{Overlaid still frames showing the coalescence of two colliding LMs. Frames shown at \SIlist{0;117;159;184;209;234;250}{\milli\second}.}
  \label{fig:coalesce}
\end{figure}

It was possible to break the SSM analogy by increasing the speed of the LMs. The speed of the LMs was calculated by measuring the distance travelled by the LM in a certain number of frames, and knowing the recording frames per second. When the collision happens at \SI{0.21}{\metre\per\second}, the LMs bounce elastically following SSM paths. However, when the speed of collision is increased to \SI{0.29}{\metre\per\second}, the two LMs coalesce. This can be seen in the video snapshots in figure~\ref{fig:coalesce}.

Growth by coalescence of LMs is a commonly observed effect \cite{Bhosale2012}. For a computing device, the physical nature of the input and output signals should be exactly the same. When two LMs coalesce, the output mass is double a single input mass. Consequently, if colliding LMs at this higher speed, a splitting device would be required to reduce the mass of the output LM. The facile splitting of a LM using a superhydrophobic treated scalpel has previously been reported \cite{Bormashenko2011scalpel}.

\begin{figure}[htbp]
\centering
\subfigure[]{\includegraphics[width=0.7\linewidth]{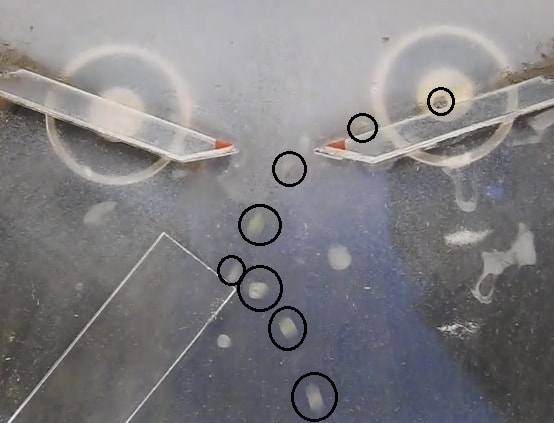}}
\subfigure[]{\includegraphics[width=0.35\linewidth]{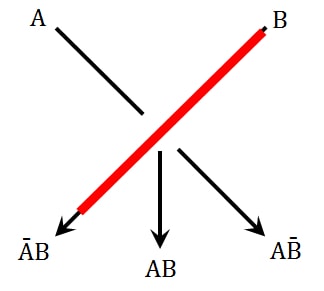}}
\subfigure[]{\includegraphics[width=0.35\linewidth]{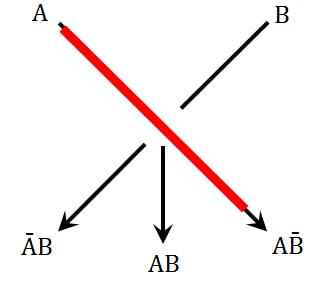}}
\subfigure[]{\includegraphics[width=0.35\linewidth]{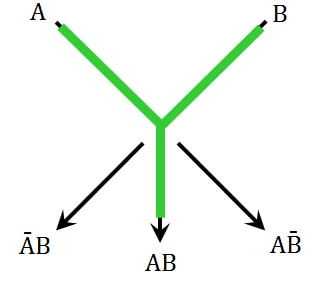}}
\subfigure[]{\includegraphics[width=0.35\linewidth]{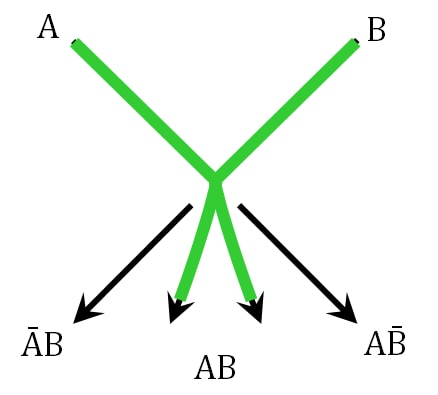}}
\subfigure[]{\includegraphics[width=0.4\linewidth]{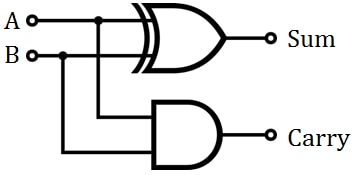}}
\caption{(a) Overlaid frames showing the successful reflection of a LM, frames are timed at \SIlist{0;234;334;375;400;434;476;517}{\ms}.
(b) \& (c) The outcomes of a single unreflected LM passing through the adder. 
(d) The outcome of the adder when two LMs collide and coalesce. 
(e) The outcome of the adder when two LMs collide according to the SSM. 
(f) The electronic representation of a 1-bit half-adder.}
  \label{fig:half-adder}
\end{figure}

By analysing the output paths of the LM collider, it becomes apparent that the gate could be modified to act as a 1-bit half-adder, with the possible outcomes demonstrated in figure \ref{fig:half-adder}. When a single LM traverses the system from the \textbf{A} or \textbf{B} channel, it finishes at the left or right extremes, the \textbf{A\={B}} or \textbf{\={A}B} path, respectively. Once the exit pathways are combined, this is analogous to the sum output on a half-adder. An initial trial confirming feasibility of this is shown in figure~\ref{fig:half-adder}(a), where a single LM enters from the right channel, crosses the gap, and is reflected to exit on the right side.

When two synchronised LMs pass through the collider, they either rebound or coalesce, according to their velocity at impact. If the LMs coalesce, then the new LM travels straight down the only \textbf{AB} path, which can be considered to be the carry output. Alternatively if the LMs rebound, as in the SSM, then there are two \textbf{AB} paths. One of these paths is then considered to be the carry output, and the other is discarded (the choice between the two \textbf{AB} paths is arbitrary in this case).

\subsection{One-Bit Full-Adder Proposed}

\begin{figure}[htbp]
  \centering
  \includegraphics[width=0.8\linewidth]{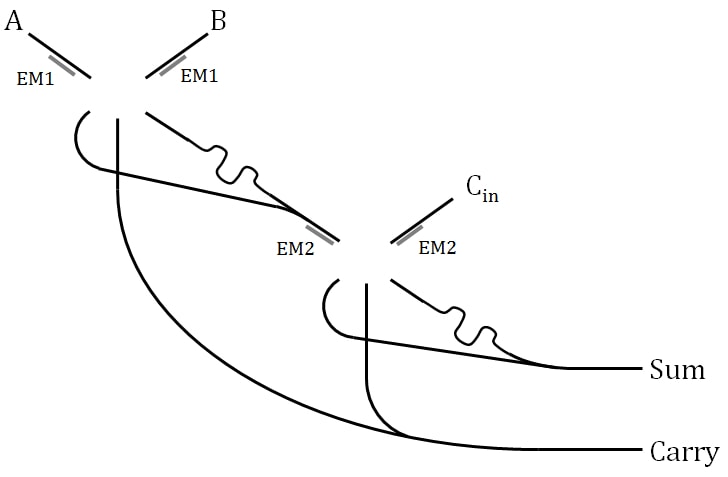}
  \caption{The general design schematic for a 1-bit full-adder, operated using liquid marbles.}
  \label{fig:full-adder1}
\end{figure}

By using this design (complete with magnetic timing control) and an intuitive {\sc xor} gate, we can adapt the model of the one-bit full-adder proposed originally for Belousov-Zhabotinsky medium \cite{adamatzky2015binary}. The {\sc xor} gate can be replaced with an {\sc or} gate without loss of logic, as there is no situation where two LMs will arrive simultaneously. The design schematic can be seen in figure~\ref{fig:full-adder1}. There are two sets of electromagnets, which cycle on-off in pairs; first EM1 releases, then shortly after EM2 releases. This maintains synchronisation between LMs across the two collision gates.

For this design iteration, we have used channels for the passage of LMs. This is a deviation from pure collision-based computing, where free-space is used as momentary `wires' on an ad hoc basis. However, in this case, we believe the use of channels to be an important intermediate step towards this goal.

\begin{figure}[htbp]
\centering
\subfigure[]{\includegraphics[width=0.494\linewidth]{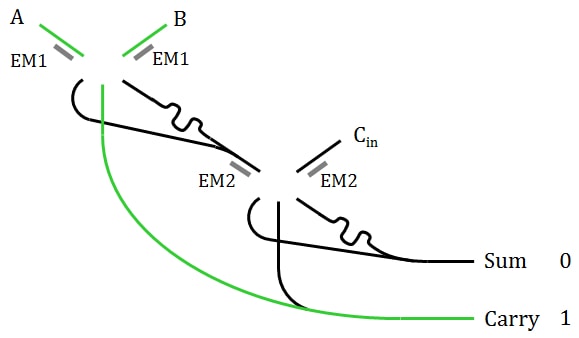}}
\subfigure[]{\includegraphics[width=0.494\linewidth]{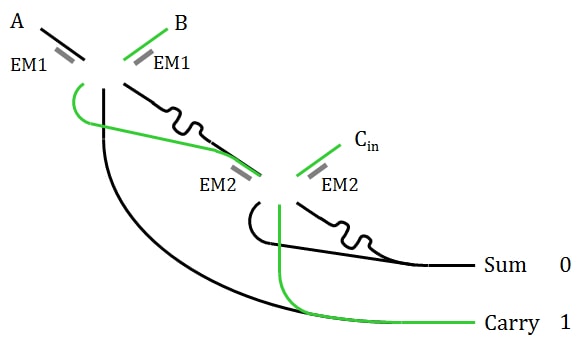}}
\subfigure[]{\includegraphics[width=0.494\linewidth]{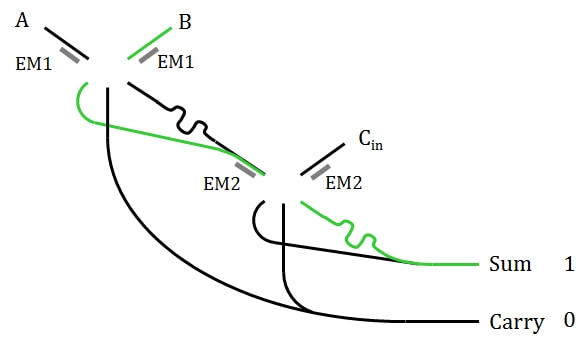}}
\subfigure[]{\includegraphics[width=0.494\linewidth]{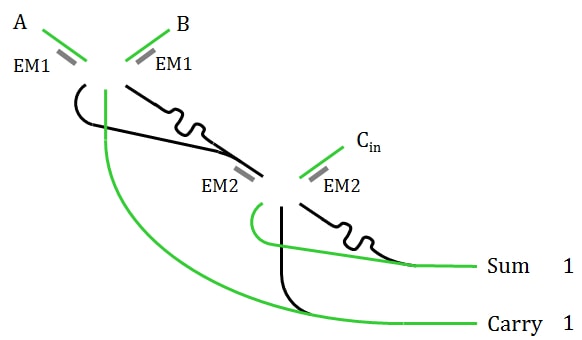}}
\caption{Example operations of the 1-bit full-adder to sum 
(a)~${1+1+0=10}$, 
(b)~${0+1+1=10}$,
(c)~${0+1+0=01}$, and 
(d)~${1+1+1=11}$.}
\label{fig:full-adder2}
\end{figure}

For the example operations visualised in figure \ref{fig:full-adder2}, if a LM travels down the \textbf{A} and \textbf{B} channel, then they will collide and travel straight to the carry output. If a LM travels down the \textbf{B} and \textbf{C\textsubscript{in}} paths, then the \textbf{B} LM crosses the first gate, before colliding with the \textbf{C\textsubscript{in}} LM and travelling straight down to join the carry output. If a single LM travels down the \textbf{B} path, it will cross the first and second gate, finishing on the sum output. If a LM travels down the \textbf{A}, \textbf{B} and \textbf{C\textsubscript{in}} paths, then \textbf{A} and \textbf{B} will collide at the first gate and go straight to the carry output, whilst the \textbf{C\textsubscript{in}} LM will cross its gate and finish on the sum output.

Based on the observations of our collision gate, we note that such a full-adder could be implemented with dimensions of approximately \SI{15 x 15}{\cm}, using LMs with a volume of approximately \SI{10}{\micro\litre}, and a LM pre-collision run length of \SI{2}{\cm}. This gate could then be cascaded as required to produce an n-bit full-adder. Empirical testing has shown that reliable and manipulable LMs can be formed down to \SI{1.0}{\micro\litre}, meaning that the device could then be scaled down appropriately. Using a single set of syringes, the automatic marble maker can up to eight LMs per needle per second. At this speed, synchronisation of the electromagnets becomes both more crucial and more tricky --- potentially requiring computer control. Initial investigations into the collision lifetime of the LM are promising, with six \SI{3}{\mm} LMs confined to a \SI{2.5 x 2.5}{\cm} space on an orbital shaker at 100\,rpm, showing no signs of wear after three hours.

\section{Conclusions}

In summary, this demonstration of a collision interaction gate represents the first computing device operated by LMs. A new automatic technique for the easy and reproducible synthesis of LMs enhances the reliability of gate operations. The novel electromagnetic synchronisation of the LM collisions was made possible by the development of a new magnetic hybrid LM, with a coating composed of nickel and UHDPE, used in conjunction with electromagnets for breaking, holding, and synchronised release of the LMs. This collision gate would operate as a 1-bit half-adder, once the sum outputs are combined. A design schematic for a 1-bit full-adder was proposed.

The use of LMs for collision-based computation has many advantages (additional degrees of freedom) over previous approaches. Due to their nature, it is possible to carry cargo in the LMs, which adds an additional dimension to the calculations. It is also possible to initiate chemical reactions within marbles by their coalescence \cite{Chen2017}. By varying the diameters of the LMs, different sizes can represent different values, and will have different relative trajectories --- removing the limitations of a binary system. Use of a magnet can remove the coating from magnetic marbles, which (if done on a superhydrophobic surface) can roll freely down a slope as droplets, before being reformed using a different coating. Compared to droplet computing \cite{Mertaniemi2012,Katsikis2015}, only a tiny portion of the circuit needs to be treated hydrophobic, making larger and more complicated circuits easier and cheaper to construct. These points, combined with LM's ability to be easily merged, levitated \cite{Chen2017}, divided \cite{Bormashenko2011scalpel}, opened/closed \cite{Zhao2010}, and easily propelled by a variety of methods make LMs a fascinating and potentially prosperous addition to the unconventional computing family.  

We envision the continued development of LM arithmetic circuits, and are currently working on producing working models of cascading standard gates.

\section*{Acknowledgements}

This research was supported by the EPSRC with grant EP/P016677/1 ``Computing with Liquid Marbles''.
The authors thank Dr Richard Mayne for his help with microscope imaging.

\section*{Supporting Information}

Video footage of the liquid marble collisions is available at \url{youtu.be/Lt1VWBtRk6E}, \url{youtu.be/isufSrhW_8M}, \url{youtu.be/sFEtVaFfxKI}, and \url{youtu.be/H8Si883lCw4}.

\section*{References}
\bibliography{Paper1}

\end{document}